\documentclass[aps,prl,preprintnumbers,superscriptaddress,showpacs,showkeys,floatfix,amsmath,amssymb,amsfonts,twocolumn]{revtex4-2}

\usepackage{graphicx}
\usepackage{multirow}
\usepackage{xparse} 
\usepackage{float}
\usepackage{xcolor}
\usepackage{makecell}
\usepackage{enumitem}
\usepackage{braket}

\NewDocumentCommand\Nf{mgg}{N\textsubscript{f}=#1\IfNoValueTF{#2}{}{+#2}\IfNoValueTF{#3}{}{+#3}}
\NewDocumentCommand\vol{mg}{#1\textsuperscript{3}\IfNoValueTF{#2}{}{×#2}}

\usepackage[normalem]{ulem}

\usepackage[colorlinks=true,linkcolor=blue,urlcolor=blue,citecolor=blue,anchorcolor=blue]{hyperref}
\usepackage[capitalise]{cleveref}
\Crefname{section}{Sec.}{Secs.}
\crefrangelabelformat{equation}{(#3#1#4--#5#2#6)}
\usepackage{slashed}

\begin{document}

\title{Spin and momentum  fraction carried by partons in the nucleon }

\author{Constantia Alexandrou} \affiliation{Department of Physics, University of Cyprus, P.O. Box 20537, 1678 Nicosia, Cyprus}\affiliation{Computation-based Science and Technology Research Center, The Cyprus Institute, 20 Kavafi Str., Nicosia 2121, Cyprus}
\author{Simone Bacchio} \affiliation{Computation-based Science and Technology Research Center, The Cyprus Institute, 20 Kavafi Str., Nicosia 2121, Cyprus}
\author{Jacob Finkenrath}\affiliation{Department of Physics, Bergische Universität Wuppertal, Gaußstraße 20, Wuppertal, 42119, Germany}
\author{Christos Iona}\affiliation{Department of Physics, University of Cyprus, P.O. Box 20537, 1678 Nicosia, Cyprus} \affiliation{Computation-based Science and Technology Research Center, The Cyprus Institute, 20 Kavafi Str., Nicosia 2121, Cyprus}
\author{Giannis Koutsou} \affiliation{Computation-based Science and Technology Research Center, The Cyprus Institute, 20 Kavafi Str., Nicosia 2121, Cyprus}
\author{Christian Kummer} \affiliation{Department of Physics, University of Cyprus, P.O. Box 20537, 1678 Nicosia, Cyprus}\affiliation{Technical University of Berlin, Berlin, Germany}
\author{Yan Li} \affiliation{Computation-based Science and Technology Research Center, The Cyprus Institute, 20 Kavafi Str., Nicosia 2121, Cyprus}
\author{Bhavna Prasad}\affiliation{Computation-based Science and Technology Research Center, The Cyprus Institute, 20 Kavafi Str., Nicosia 2121, Cyprus}
\author{Gregoris Spanoudes}\affiliation{Department of Physics, University of Cyprus, P.O. Box 20537, 1678 Nicosia, Cyprus}

\date{\today}

\begin{abstract}
We determine the momentum fraction and angular momentum  carried by quarks and gluons in the proton in lattice QCD. We use four ensembles simulated with up, down, strange and charm quarks with their masses tuned to  their physical values. 
These  ensembles have similar physical volume and different lattice spacings  allowing us to take the continuum limit directly at the physical pion mass point. 
We extract the quark and gluon momentum fractions and total angular momentum in the continuum limit as well as the intrinsic quark spin and orbital angular momentum contributions to the proton spin. We find the total momentum fraction $\braket{x}_N = 0.995(60)(29)$ and the total spin $J_N = 0.507(43)(65)$, showing that
both the momentum and spin sum rules are satisfied. We compare our results to those extracted from phenomenological analyses.

\centerline{\includegraphics[width=0.2\linewidth]{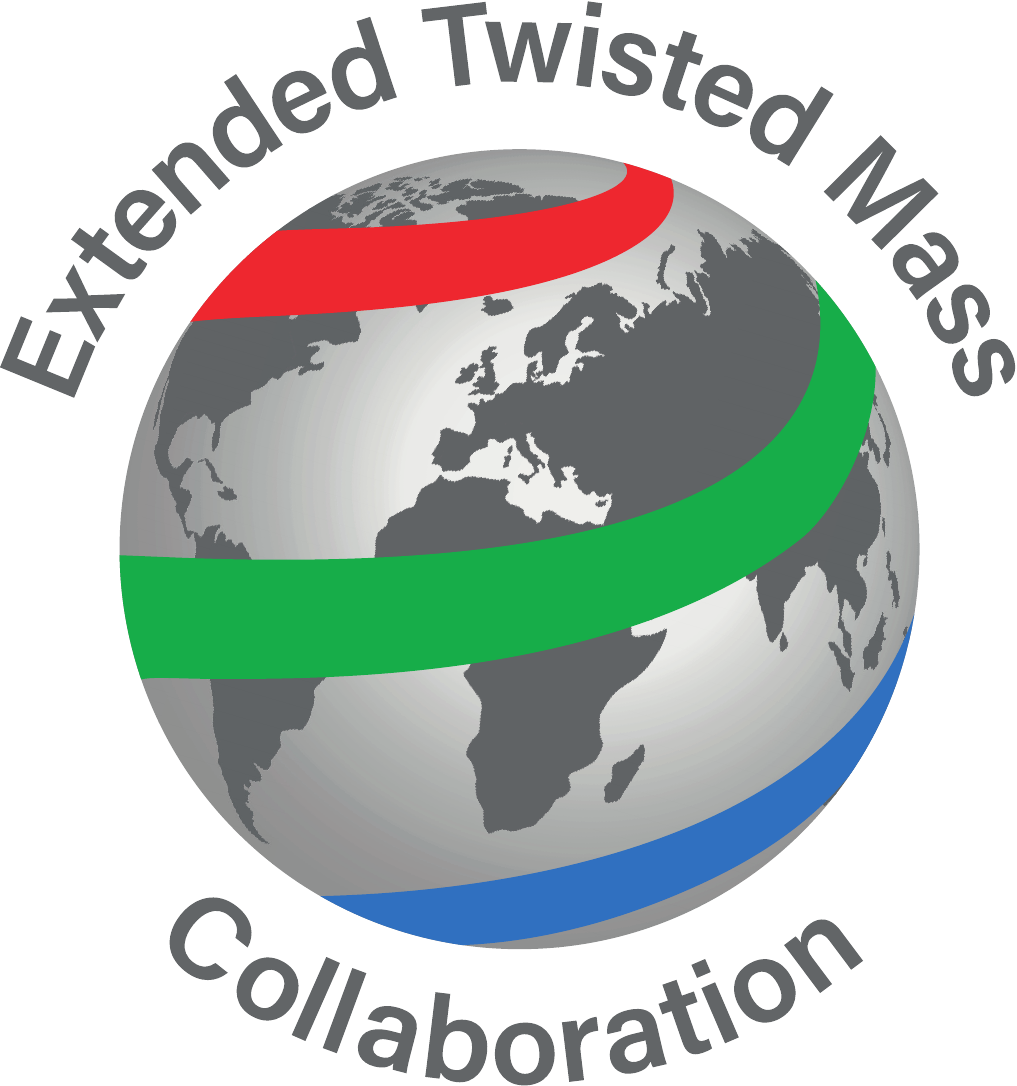}}
\end{abstract}

\maketitle

\paragraph{Introduction.}
Understanding how the structure of the proton arises from the underlying theory of the strong interactions is the focus of several current and future experiments~\cite{Aidala:2012mv,STAR:2014wox,Accardi:2012qut} via 
deep inelastic scattering (DIS), semi-inclusive DIS, and polarized proton-proton collisions.  One fundamental question, raised as early as the late 1980s by the European Muon Collaboration (EMC) at CERN, is how the momentum and spin of the nucleon are distributed among quarks and gluons. The first such  precise measurement, using polarized deep inelastic muon–proton scattering, revealed  the surprising result that the spin of the   valence  quarks accounts for only a small part of the  spin of the proton~\cite{Ashman:1988hv,Ashman:1989ig}, far less compared to what was expected at the time, and which resulted in the so-called  {\it proton spin puzzle}. The EMC result triggered decades of theoretical and experimental work, motivating
polarized DIS experiments, such as SMC, HERMES, and COMPASS~\cite{SpinMuon:1998eqa,HERMES:2006jyl,COMPASS:2016xvm},
polarized proton–proton programs at the Relativistic Heavy Ion Collider,
global parton distribution function (PDF) analyses, and modern nucleon structure techniques in lattice Quantum Chromodynamics (QCD). Seeking a full understanding of how proton structure emerges from QCD persists as one of the major goals of nucleon and particle physics and a main motivation for building  the  Electron-Ion Collider~\cite{STAR:2014wox}. A recent overview on the subject is given in Ref.~\cite{Ji:2020ena}.

In this work, we use modern methods in lattice QCD to  determine the
decomposition of the spin of the proton into the contributions from
its constituent quarks and gluons, addressing the proton spin puzzle  from first principles. A by-product of this calculation is the determination of  the momentum fraction carried by quarks and gluons. Although over the past five years~\cite{Alexandrou:2026soz,Alexandrou:2020sml,Hackett:2023rif}, there has been progress in the determination of the spin content of the proton using lattice QCD, this work is the first complete quark flavor and gluon decomposition of the proton spin in lattice QCD using only simulations with quark masses fixed to their physical values and taking the continuum limit without any chiral extrapolations.

\paragraph{Matrix element of the nucleon energy-momentum tensor.}
The quark and gluon contributions to the nucleon momentum fraction and angular momentum are encoded in matrix elements of the QCD  energy-momentum tensor (EMT). 
We consider the traceless symmetric gauge-invariant EMT, $T^{\mu\nu}$, decomposed into quark and gluon contributions as 
\begin{align}\label{eq:EMT}
    T^{\mu\nu}=T^{\mu\nu}_q+T^{\mu\nu}_g =  \bar{\psi}i\gamma^{\{\mu} \overleftrightarrow{D}^{\nu\}} \psi  + F^{\rho\{\mu} F^{\nu\}}_{\ \rho} \,,
\end{align}
where $F^{\mu\nu}$ is the gluon field-strength tensor and the notation $\{\cdots\}$ means symmetrization over the indices in the curly brackets and subtraction of the trace. The symmetrized  covariant derivative $\overleftrightarrow{D}$ is defined as $(\overrightarrow{D}-\overleftarrow{D})/2$.

Going to Euclidean space and  following the convention of Ref.~\cite{Hagler:2003jd}, the nucleon matrix element of  EMT is parameterized in terms of the gravitational form factors (GFFs) as
\begin{align}\label{eq:me2ff}
& \braket{N(p^\prime, s^\prime) | T^{\mu\nu}_{q,g} | N(p, s) } = 
\bar{u}_N(p^\prime, s^\prime) \Bigg[
iA^{q,g}_{20}(Q^2)\, \gamma^{\{\mu} P^{\nu\}} \nonumber\\
&+B^{q,g}_{20}(Q^2)\, \frac{P^{\{\mu} \sigma^{\nu\}\rho} q_\rho}{2m_N} 
+\, C^{q,g}_{20}(Q^2)\, \frac{q^{\{\mu} q^{\nu\}}}{m_N}
\Bigg] u_N(p, s),
\end{align}
where $N(p,s)$ denotes the nucleon state with initial (final) momentum $p$ ($p^\prime$) and spin $s$ ($s^\prime$), $u_N$ is the Euclidean nucleon spinor, $m_N$ is the nucleon mass, $P=(p^\prime+p)/2$, $q=p-p^\prime$, and $Q^2=-q^2$.

In the forward limit, the form factor $A^{q,g}_{20}(0)$ yields the momentum fraction carried by quarks and gluons,
\begin{align}
\braket{x}_{q,g} = A^{q,g}_{20}(0),
\end{align}
which should satisfy the momentum sum rule $\braket{x}_q+\braket{x}_g=1$. The total spin carried by quarks and gluons is obtained via \cite{Ji:1996ek}
\begin{align}
J_{q,g} = \frac{1}{2}\left[A^{q,g}_{20}(0) + B^{q,g}_{20}(0)\right]
\end{align}
and the spin sum is $J_q+J_g = 1/2$. In this paper, we check the momentum and spin sums by computing {\it ab initio} all components.

\paragraph{Lattice setup.}
We use twisted mass clover-improved  fermion ensembles simulated with two degenerate up and down quarks, a strange and a charm quark  with quark masses (\Nf{2}{1}{1}) tuned to approximately
their physical values. This formalism allows for automatic
${\cal{O}}(a)$ improvement without requiring any further improvement
of the operators~\cite{Frezzotti:2000nk,
Frezzotti:2003ni}. A summary of the parameters of the ensembles is
provided in Table~\ref{tab:ens}. 
\begin{table}[h]
	\caption{Parameters of the four \Nf{2}{1}{1} ensembles analyzed in
		this work. From the leftmost to the rightmost columns, we provide
		the name of the ensemble and its abbreviation in parenthesis, the lattice volume,  the lattice spacing, the pion
		mass, $m_\pi$, and the value of $m_\pi L$. Lattice spacings and pion
		masses are taken from
Refs.~\cite{ExtendedTwistedMass:2022jpw,ExtendedTwistedMass:2024nyi}.}
	\label{tab:ens}
	\centering
	\begin{tabular}{ccccccc}
		\hline\hline
		Ensemble   (short name)          & $(\frac{L}{a})^3{\times}(\frac{T}{a})$ &  \makecell[c]{$a$                   \\$[$fm$]$} & \makecell[c]{$m_\pi$\\ $[$MeV$]$}  & $m_\pi L$ \\
		\hline
		\texttt{cB211.072.64} (B64) & $64^3 {\times} 128$                      & 0.07948(11)      & 140.2(2) & 3.62 \\
		\texttt{cC211.060.80} (C80) & $80^3 {\times} 160$                      & 0.06819(14)      & 136.7(2) & 3.78 \\
		\texttt{cD211.054.96} (D96) & $96^3 {\times} 192$                       & 0.056850(90)      & 140.8(2) & 3.90 \\
		\texttt{cE211.044.112} (E112)& $112^3 {\times} 224$                      & 0.04892(11)      & 136.5(2) & 3.79 \\
		\hline
	\end{tabular}
\end{table}

In order to determine the nucleon matrix elements of  EMT, we compute  nucleon two- and three-point functions and  employ  smearing to enhance overlap with the nucleon ground state.
For the gauge links entering $T^{\mu\nu}_{g}$, we apply stout smearing~\cite{Morningstar:2003gk} to improve the signal-to-noise ratio, as described in Ref.~\cite{Alexandrou:2020sml}. 
The nucleon three-point functions receive both connected and disconnected contributions for which we employed improved techniques~\cite{Alexandrou:2020sml}, as described in more detail in Ref.~\cite{Alexandrou:2026oks}, where we also give the statistics.

\paragraph{Non-perturbative renormalization.} 
The matrix elements of $T^{\mu\nu}_{g}$ and 
$T^{\mu\nu}_{q}$ are renormalized non-perturbatively 
in the RI$'$-MOM scheme~\cite{Martinelli:1994ty}, followed by 
a perturbative conversion to the $\overline{\rm MS}$ scheme 
at the scale  of $\bar{\mu} = 2$ GeV. We consider the flavor non-singlet ($ns$) combinations $T^{\mu\nu;ns}_{q} \in \{T^{\mu\nu}_{u-d}$, $T^{\mu\nu}_{u+d-2s}$, $T^{\mu\nu}_{u+d+s-3c}\}$, as well as the flavor-singlet ($s$) combination $T^{\mu\nu;s}_{q} \equiv T^{\mu\nu}_{u+d+s+c}$. The non-singlet operators are renormalized multiplicatively with a common renormalization factor $Z_{qq}$ as $(T^{\mu\nu;ns}_{q})^{\mathrm{R}} = Z_{qq} \, T^{\mu\nu;ns}_{q}$.

In contrast, the quark singlet and gluon EMT components mix under renormalization
according to:
\begin{equation}
  \begin{pmatrix}
    (T^{\mu\nu;s}_{q})^{\mathrm{R}}\\
    (T^{\mu\nu}_{g})^{\mathrm{R}}
  \end{pmatrix}
  =
  \begin{pmatrix}
    Z_{qq}^s & Z_{qg}\\
    Z_{gq} & Z_{gg} \\
  \end{pmatrix}
  \begin{pmatrix}
    T^{\mu\nu;s}_{q} \\
    T^{\mu\nu}_{g}
  \end{pmatrix}.
  \label{Eq:mixing_matrix}
\end{equation}

A detailed description of the non-perturbative computation of all renormalization coefficients is provided in Ref.~\cite{Alexandrou:2026oks}. Compared to our previous study on the spin decomposition~\cite{Alexandrou:2020sml}, where only the coarser ensemble B64 is used, several improvements are implemented to reduce systematic uncertainties. In the present analysis, we determine both diagonal and off-diagonal elements of the mixing matrix in Eq.~\eqref{Eq:mixing_matrix} fully non-perturbatively and extend the study to the three finer lattice ensembles. 
We further introduce improved renormalization conditions that reduce rotational breaking effects by selecting momentum values close to the body-diagonal direction for all vertex functions. Furthermore, we suppress mixing with gauge non-invariant operators in the RI$'$-MOM vertex functions, which has not been considered in previous renormalization studies. This is achieved by imposing an enlarged set of renormalization conditions and by explicitly computing additional components entering the intermediate stages of the analysis.

\paragraph{Extraction of the momentum fraction and angular momentum.}
To extract the GFFs, we construct an overconstrained system of linear equations and solve it using the its singular value decomposition~\cite{Alexandrou:2020sml}.  Details on the analysis are given in Ref.~\cite{Alexandrou:2026oks}. 

The GFF $A_{20}(Q^2)$, can be  obtained directly at $Q^2=0$ yielding the momentum fraction, namely $\langle x\rangle = A_{20}(0)$. One can also examine the $Q^2$-dependence of $A_{20}(Q^2)$ using data for finite values of $Q^2$ and extrapolating to $Q^2=0$, thus providing a cross-check. For $B_{20}(0)$, the kinematic factor multiplying the relevant matrix element means that it cannot be determined directly at $Q^2=0$ and thus an extrapolation is necessary.  

Once the values of $A_{20}(0)$ and  $B_{20}(0)$ are determined and renormalized, we extrapolate to the continuum limit, which is carried out   using both constant and linear fits in $a^2$ and  averaged weighted by their Akaike information criterion (AIC)~\cite{Jay:2020jkz,Neil:2022joj}. 
The extrapolations are performed for the isoscalar and the three isovector flavor combinations  as well as for the gluon contribution. 
The individual  and the total quark momentum fractions are then reconstructed through linear combinations of the continuum extrapolated results. 
 One can see that, while the dependence on the lattice spacing is mild, taking the continuum limit is important for obtained results with controlled systematics.

\begin{figure}[!ht]
	\centering
    \includegraphics[width=\columnwidth]{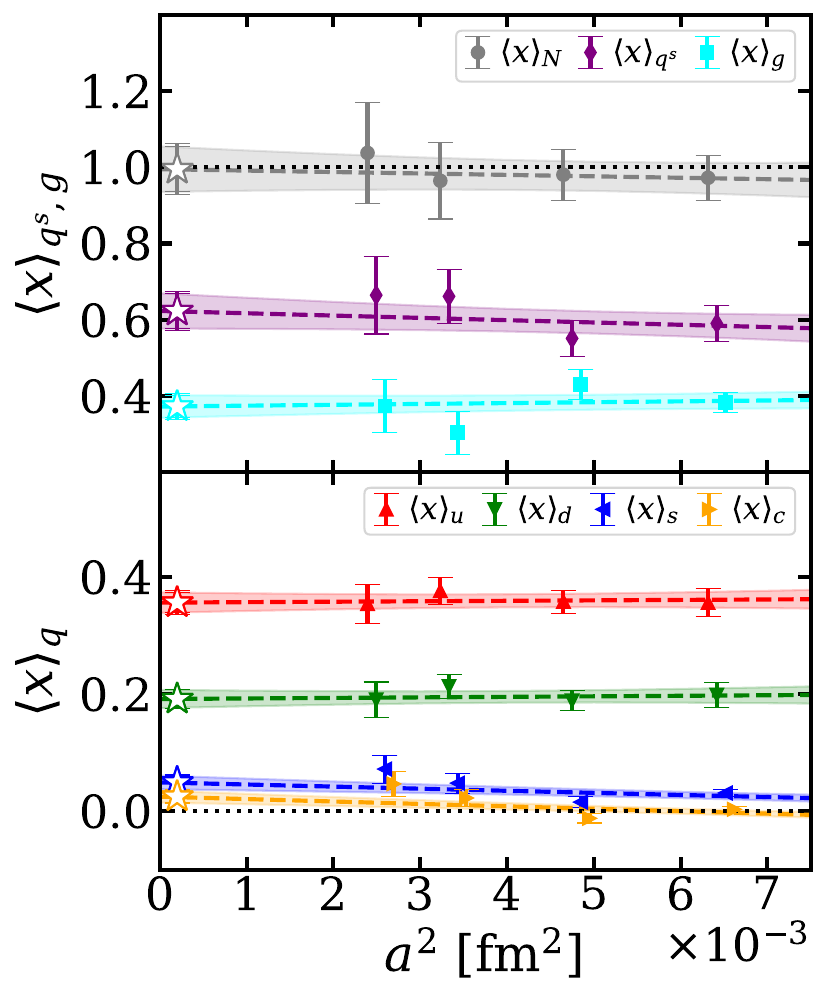}
    \caption{
    Continuum extrapolation of the proton momentum fractions.
    Top: Results on the total quark (purple diamonds) and gluon (cyan squares) momentum fractions and their sum (grey circles). Bottom: Results on the  up (red upwards pointing triangles), down (green downwards pointing triangles), strange (blue left pointing triangles), and charm (yellow right pointing triangles) quark momentum fractions. 
    The stars at $a=0$ show continuum-limit results. For the continuum results, the inner error bars show the statistical uncertainties while the outer error bars show statistical and systematic uncertainties added in quadrature. In most cases, systematic errors are too small to discern.  
    }
    \label{fig:x_cont}
\end{figure}

\begin{figure}[h!]
    \centering
    \includegraphics[width=\columnwidth]{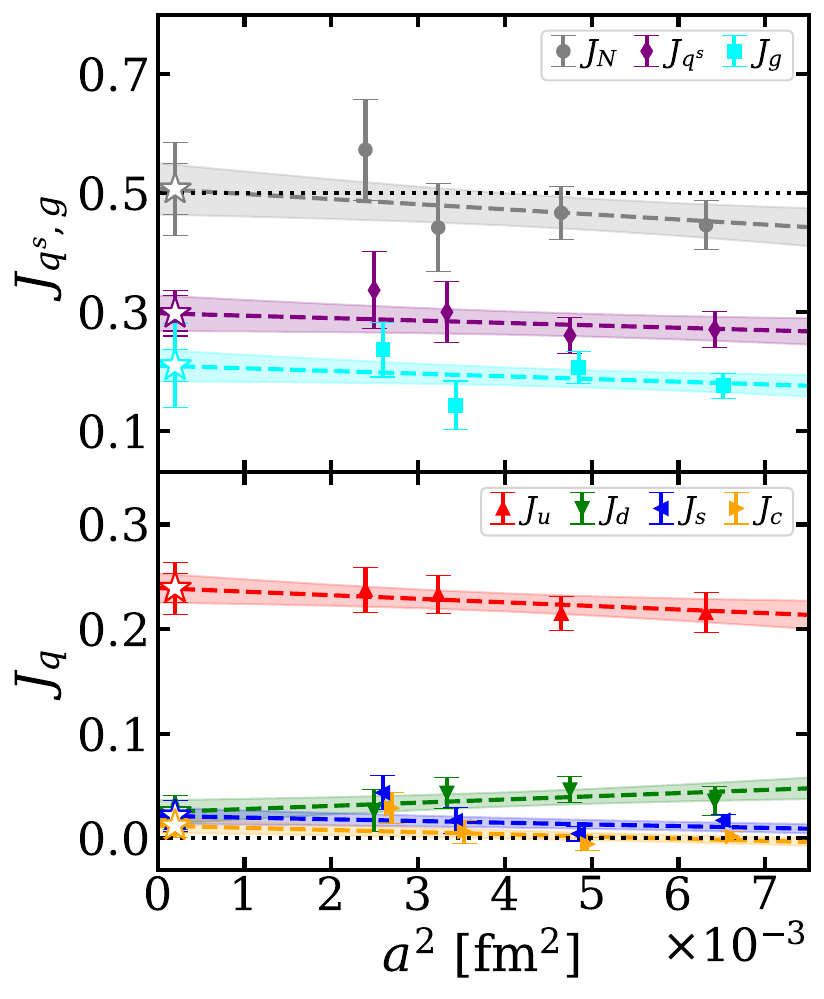}
    \caption{Results on the proton  angular momentum  fractions for the four ensembles analyzed and  their continuum limit. The notation is the same as that in  Fig.~\ref{fig:x_cont}. }
    \label{fig:J_cont}
\end{figure}

\begin{figure}[h!]
    \centering
    \includegraphics[width=\columnwidth]{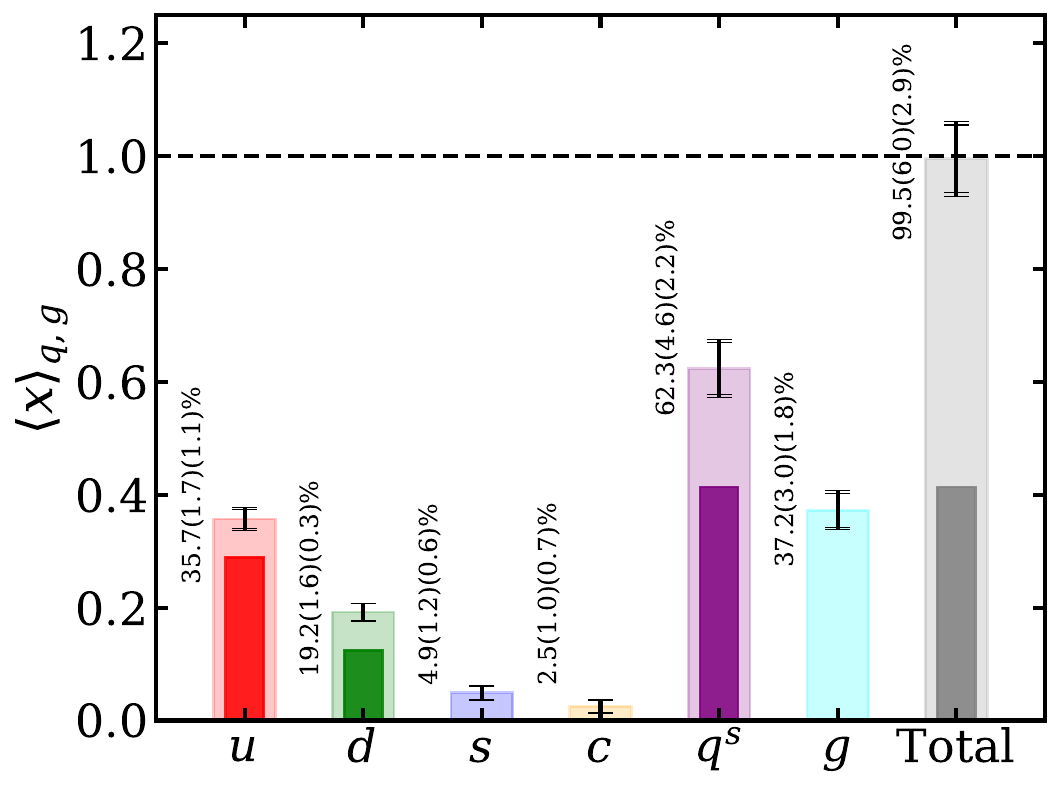}
    \caption{
    Continuum-extrapolated  quark and gluon momentum fractions in the proton at the physical pion mass. 
    Inner bars show connected quark contributions  and outer  the total.
    }
    \label{fig:bar_x}
\end{figure}

\begin{figure}[h!]
    \centering
    \includegraphics[width=\columnwidth]{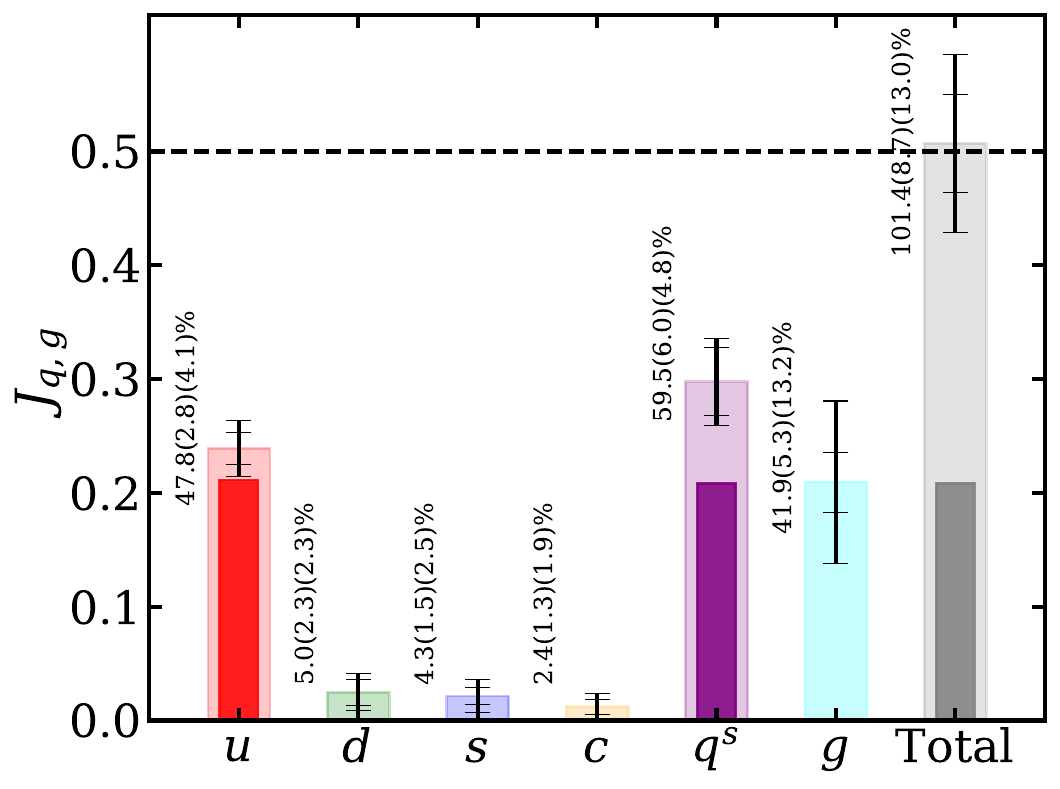}
    \caption{
    Continuum-extrapolated results  for the  proton total angular momentum. The notation is the same as in Fig.~\ref{fig:bar_x}.
    }
    \label{fig:bar_J}
\end{figure}
\begin{figure}[h!]
    \centering
    \includegraphics[width=\columnwidth]{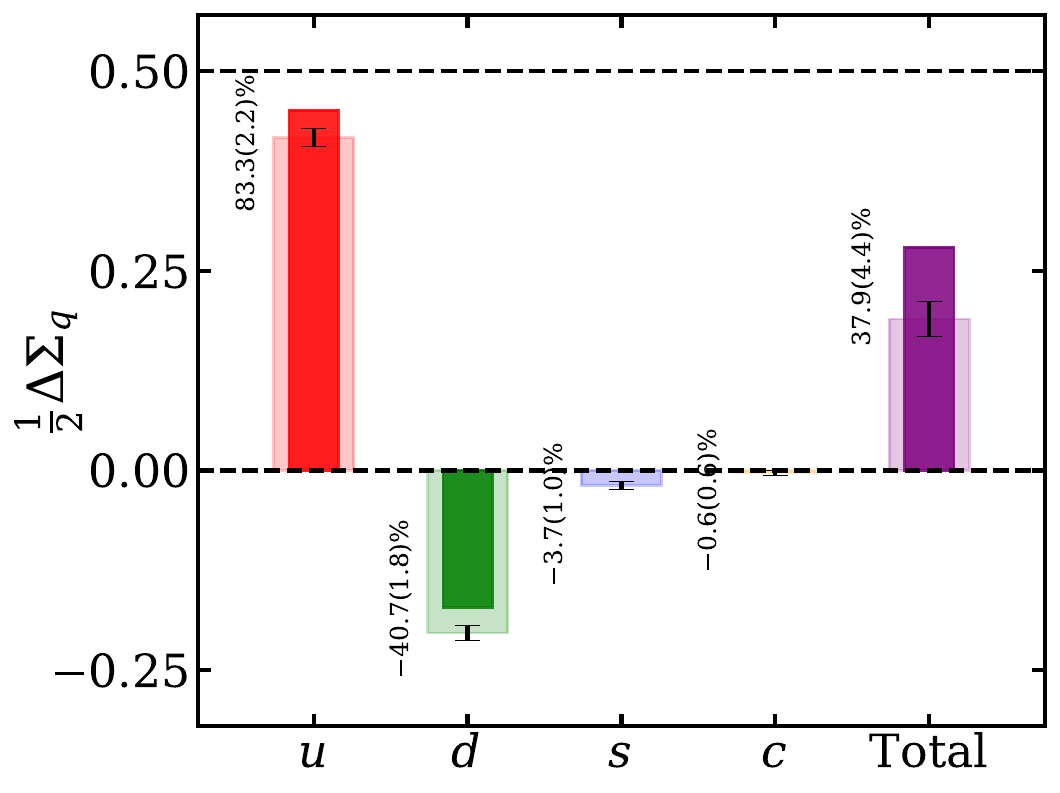}\\
     \includegraphics[width=\columnwidth]{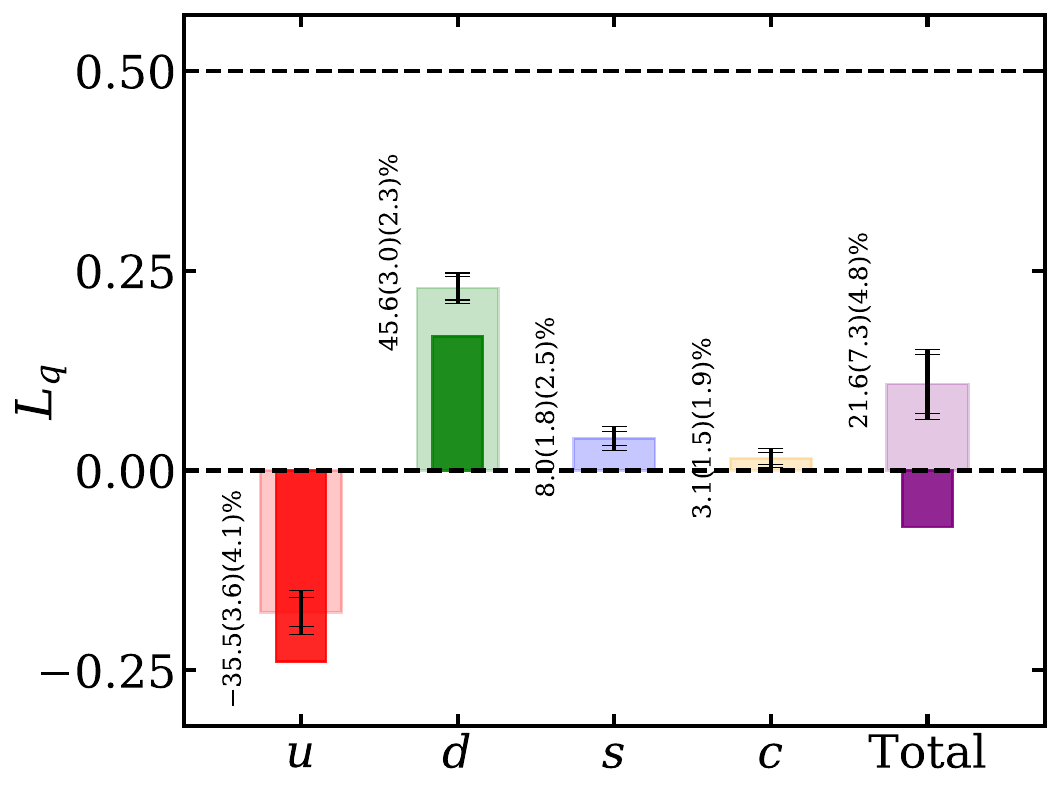}
    \caption{
   Continuum extrapolated results  for the intrinsic quark spin (upper panel) and for the orbital angular momentum (lower panel) in the proton. The notation is the same as in Fig.~\ref{fig:bar_x}.
    }
    \label{fig:bar_DeltaSigma_L}
\end{figure}

\begin{table*}[!ht]
    \caption{
    Results for the proton average momentum fraction $\braket{x}$, total angular momentum $J$, intrinsic quark spin $\frac{1}{2}\Delta\Sigma$, and orbital angular momentum $L$ in the $\overline{\rm MS}$ scheme at 2 GeV. Results are given separately for the up ($u$), down ($d$), strange ($s$), charm ($c$), and gluon ($g$) contributions. The sum over quarks and gluons is given in the last row. For $\braket{x}$, $J$, and $L$, computed in this work,  the first error combines statistical error with the AIC model average error of the continuum extrapolation, and the second error combines the systematic errors due to the excited-state analysis, the ranges used in the  $Q^2$ extrapolation and the gluon-stout-smearing analysis, see Ref.~\cite{Alexandrou:2026oks} for details. 
    For $\frac{1}{2}\Delta \Sigma_q$ taken form Ref.~\cite{christos}, the error corresponds to our first one.
    }\label{tab:val}
    \centering
    \renewcommand\arraystretch{1.2}
    \begin{ruledtabular}
\begin{tabular}{cccccc}
 & $\braket{x}$ & $J$ & $\frac{1}{2}\Delta\Sigma$ & $L$ \\
\hline
u & 0.357(17)(11) & 0.239(14)(21) & 0.417(11) & -0.177(18)(21) \\
d & 0.192(16)(3) & 0.025(11)(12) & -0.2033(89) & 0.228(15)(12) \\
s & 0.049(12)(6) & 0.0215(76)(123) & -0.0187(51) & 0.0401(92)(123) \\
c & 0.025(10)(7) & 0.0121(66)(95) & -0.0032(32) & 0.0153(73)(95) \\
g & 0.372(30)(18) & 0.209(26)(66) &  &  \\
Sum & 0.995(60)(29) & 0.507(43)(65) & 0.190(22) & 0.108(37)(24) \\
\end{tabular}
    \end{ruledtabular}
\end{table*}
\paragraph{Results and discussion.}
Our results for the momentum fraction for  each  parton  are shown in Fig.~\ref{fig:x_cont}. We observe that the up quark carries about twice the momentum fraction of the  down quark, as expected from the valence structure of the proton. After taking the corresponding continuum  limit  of $B_{20}(0)$, we show in Fig.~\ref{fig:J_cont} the results for the angular momentum $J_{q,g}$, including the flavor quark decomposition. As can be seen, the major contributions come from the gluon and the up quark, the value of which is enhanced  due to the positive contribution of  $B^u_{20}(0)$. The down quark value, $J_d$, is decreased due to the  negative contribution of $B^d_{20}(0)$.

The momentum fractions and total angular momentum contributions from each parton in the continuum limit are shown in Figs.~\ref{fig:bar_x} and~\ref{fig:bar_J}, respectively. 
  The intrinsic quark spin  $\frac{1}{2}\Delta\Sigma_q=\frac{1}{2}g_A^q$, can be obtained from  the axial charges of Ref.~\cite{christos},  which are updates of the results from Refs.~\cite{Alexandrou:2024ozj,Alexandrou:2026soz}.
   In addition, the quark orbital angular momentum can be extracted using  $L_q=J_q-\frac{1}{2}\Delta\Sigma_q$. Both results are  shown in \cref{fig:bar_DeltaSigma_L}.
The corresponding  values   are collected in \cref{tab:val}. 

The up quark provides a significant contribution to the intrinsic spin. The down quark contributes with opposite sign of about half the magnitude, again consistent with the quark content of the proton. The strange contribution is small and negative, while the charm contribution is consistent with zero. 
For the orbital angular momentum $L$, interestingly and unlike for the intrinsic spin, both the up and down quarks contribute equally  and with opposite signs. The large positive down quark contribution $L_d$ largely compensates the down quark negative $\frac{1}{2}\Delta \Sigma_d$, yielding a comparatively small positive value for $J_d$.

\begin{figure*}[h!]
    \centering
    \includegraphics[width=\textwidth]{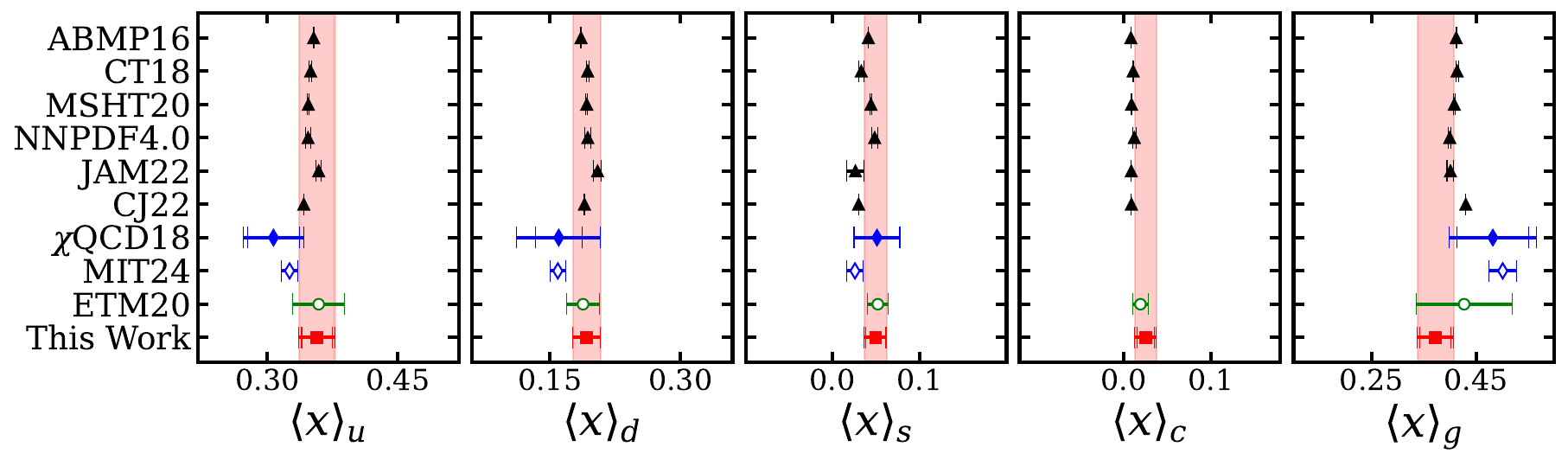}\\
    \includegraphics[width=\textwidth]{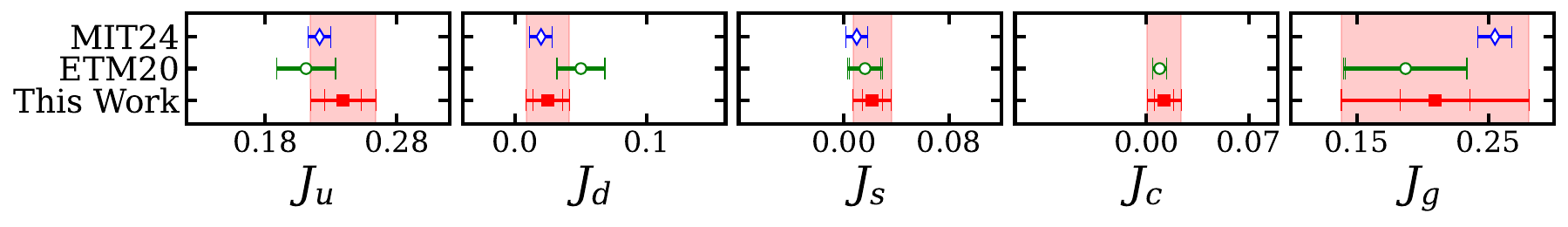}
    \caption{
    Comparison of our results for quark and gluon momentum fractions $\braket{x}_{q,g}$ (upper panels) and spin contributions $J_{q,g}$ (lower panels) with other lattice QCD results and phenomenology.
    Columns show, from left to right, the $u$, $d$, $s$, and $c$ quark contributions, followed by the gluon contribution.
    Red squares denote this work, green circles ETMC20~\cite{Alexandrou:2020sml} obtained using the B64  ensemble only, open blue diamonds are from Ref.~\cite{Hackett:2023rif}, obtained using one ensemble with $m_\pi\approx170$ MeV and $a\approx0.091$ fm, and filled blue diamonds are from $\chi$QCD18~\cite{Yang:2018nqn}.
    Open symbols indicate results without a continuum extrapolation.
    Black triangles show global PDF determinations, in ascending order from CJ22~\cite{Accardi:2023gyr}, JAM22~\cite{Cocuzza:2022jye}, NNPDF4.0~\cite{NNPDF:2021njg}, MSHT20~\cite{Bailey:2020ooq}, CT18~\cite{Hou:2019efy}, and ABMP16~\cite{Alekhin:2017kpj}.
    }
    \label{fig:compare_avgx_J}
\end{figure*}

\begin{figure*}[h!]
    \centering
    \includegraphics[width=\textwidth]{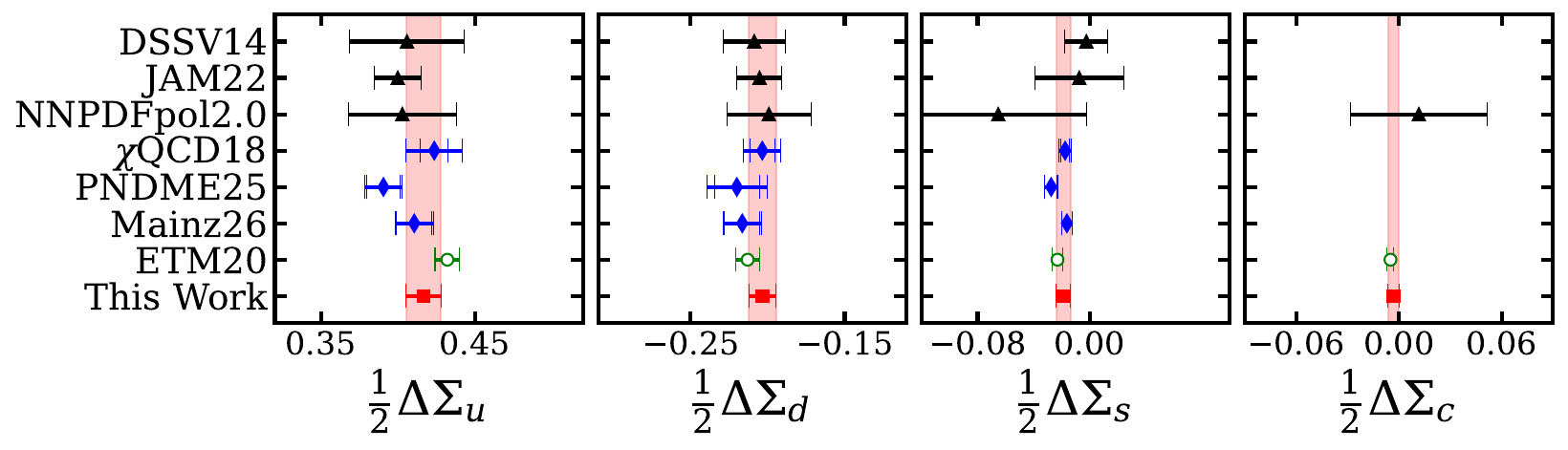}\\
    \includegraphics[width=\textwidth]{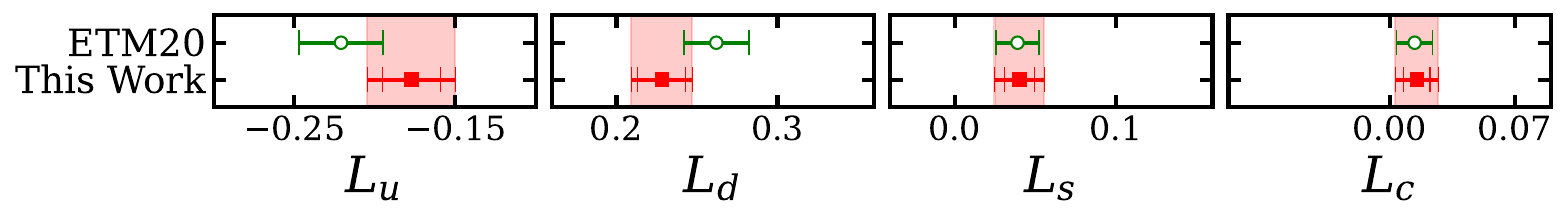}
    \caption{
    Comparison of our  results for the quark intrinsic spin contributions, $\Delta\Sigma_q/2$ (upper panels), and orbital angular momenta, $L_q$ (lower panels), with other lattice QCD results and phenomenology.
    Columns show, from left to right, the $u$, $d$, $s$, and $c$ quark contributions.
    Red squares denote this work, green circles ETMC20~\cite{Alexandrou:2020sml}, and blue symbols lattice results from PNDME25~\cite{Park:2025rxi}, Mainz26~\cite{Barone:2026uyx,Djukanovic:2024krw}, and $\chi$QCD18~\cite{Liang:2018pis}.
    Open symbols indicate results without a continuum extrapolation.
    Black triangles show phenomenological determinations in ascending order from NNPDFpol2.0~\cite{Cruz-Martinez:2025ahf},  JAM22~\cite{Cocuzza:2022jye}, and DSSV14~\cite{DeFlorian:2019xxt}.
    For the phenomenological results, $\Delta\Sigma_q=g_A^q$ is obtained by integrating the polarized PDFs over $x\in[10^{-3},1]$, since the small-$x$ region is poorly constrained by current experimental data.
    }
    \label{fig:compare_DS2_L}
\end{figure*}

In what follows, we compare our continuum results  with recent  lattice QCD calculations and phenomenological determinations. In \cref{fig:compare_avgx_J}, we show the momentum  and angular momentum fractions for each quark flavor and for the gluon. Our results for the momentum fractions are in agreement with those from phenomenological analyses. They are also in agreement with the  results from the $\chi$QCD collaboration~\cite{Yang:2018nqn}, which are the only ones where a continuum extrapolation was carried out.  The recent analysis in Ref.~\cite{Hackett:2023rif} is done using only one gauge ensemble with $m_\pi\approx170$~MeV and $a\approx0.091$~fm. Their results are in agreement with our previous analysis using the coarsest B64 ensemble~\cite{Alexandrou:2020sml}.

Our results for the proton spin  are the only ones in the continuum limit. They are in good agreement with our previous lattice QCD results using the B64 gauge ensemble, as well as with the results of Ref.~\cite{Hackett:2023rif}. The larger error on our value of $J_g$ is due to taking into account  systematic effects mostly due to the excited state analysis of the disconnected.

In Fig.~\ref{fig:compare_DS2_L}, we show a comparison of results 
 for the  intrinsic quark spin and orbital angular momentum. The results on $\frac{1}{2}\Delta \Sigma_q$ are in  agreement  with  global analyses of PDF, as well as with other lattice QCD determinations. Notably, we achieve a better precision as compared to phenomenological results, including  a  non-zero negative strange quark contribution. For $L_q$, where  we can only compare to our previous results using the B64 ensemble, we find agreement with our current results including  systematics uncertainties. 

\paragraph{Conclusions.}
We present the first determination  of the quark and gluon contributions to the proton momentum fraction and angular momentum  in the proton at the continuum limit using only gauge ensembles with physical quark masses.  All renormalization functions, including the quark singlet and its mixing with the gluon, are  determined fully non-perturbatively. 
Using four lattice spacings enables the control of lattice discretization effects through a continuum extrapolation. We find $\braket{x}_N = 0.995(60)(29)$ for the sum of the momentum fractions and $J_N = 0.507(43)(65)$ for the spin sum, where the first error is the statistical error and the second systematic error. 
The results in  Figs.~\ref{fig:bar_x} and \ref{fig:bar_J}, show that valence  contributions  account for only about half of the total momentum fraction and spin of the proton, indicating sizeable sea quark and gluon  contributions.
The gluon contribution to both the momentum and spin sums is approximately 40\%, while the sea-quark contribution amounts to about 10\%.   We thus confirm that about half of the spin of the proton comes from non-valence contributions, in agreement with the EMC result, resolving the long-standing puzzle of the spin of the proton. We also find a small but statistically non-zero strange quark contribution to all quantities. Interestingly, we also find a small non-zero contribution of the charm quark to the momentum fraction, indicating charm quark effect in the nucleon.

Our results for the momentum fractions and intrinsic quark spin are in agreement with phenomenological determinations from global parton distribution functions analyses.

The results of this work provide the first  quantitatively complete parton decomposition of the proton momentum fraction and angular momentum, and quark intrinsic spin and orbital angular momentum  directly from QCD. 

\paragraph{Acknowledgments.}
We would like to thank all members of the Extended Twisted Mass Collaboration for a very enjoyable cooperation. C.A., S.B., C.I., G.K., Y.L. and G.S.  acknowledge partial support from the projects
Baryon8 (POSTDOC/0524/0001), MuonHVP (EXCELLENCE/0524/0017), PulseQCD
(EXCELLENCE/0524/0269), DeNuTra (EXCELLENCE/0524/0455), IMAGE-N
(EXCELLENCE/0524/0459), HyperON (VISION ERC-PATH 2/0524/0001),
StrongILA (EXCELLENCE/0524/0001), RobustSigma (EXCELLENCE/0925/0325) and partonWF (VISION ERC/0525/0010)
co-financed by the European Regional Development Fund and the Republic
of Cyprus through the Research and Innovation Foundation within the framework of the Cohesion Policy Programme ``THALIA 2021-2027''.
This project received funding from the European Research Council (ERC) via the project ”LEEX” grant agreement 101170304. Funded by the European Union. Views and opinions expressed are however those of the author(s) only and do not necessarily reflect those of the European Union or the European Research Council Executive Agency (ERCEA). Neither the European Union nor the ERCEA can be held responsible for them.
C.K. is supported by European Union’s HORIZON MSCA Doctoral Networks programme, under Grant Agreement No. 101072344, project AQTIVATE (Advanced computing, QuanTum algorIthms and data-driVen Approaches for science, Technology and Engineering).
We acknowledge the Swiss National Supercomputing Centre (CSCS) access to Alps through the Chronos programme under project ID CH15
and access to the LUMI supercomputer through the Chronos programme under project IDs CH17-CSCS-CYP
. 

We gratefully acknowledge CINECA and the EuroHPC Joint Undertaking for granting access to the Leonardo Supercomputer. Computing time on Leonardo Booster was allocated through the Extreme Scale Access Call (grant EHPC-EXT-2024E01-027), and additional GPU resources were provided under the INFN-LQCD123 initiative. 
The authors gratefully acknowledge the Gauss Centre for Supercomputing e.V. (www.gauss-centre.eu) for funding this project by providing computing time through the John von Neumann Institute for Computing (NIC) on the GCS Supercomputers JUWELS~\cite{JUWELS}, JUWELS Booster~\cite{JUWELS-BOOSTER} and JUPITER Booster at the J\"ulich Supercomputing Centre (JSC).
The authors also acknowledge the Texas Advanced Computing Center (TACC) at University of Texas at Austin for providing HPC resources.

\bibliography{refs}

\appendix

\end{document}